\documentclass[a4paper,12pt]{article}
\usepackage{latexsym}
\usepackage{exscale}
\usepackage{graphicx}
\usepackage{amsmath}
\usepackage{amssymb}
\renewcommand{\vec}{\mathbf}
\usepackage[germanpar]{anysize}
\marginsize{2.5cm}{2.5cm}{2.0cm}{2.0cm}
\usepackage{natbib}
\bibpunct{(}{)}{,}{a}{}{,}
\renewcommand{\cite}{\citep}
\begin{document}
%
%
\title{Reorientation phase transitions in thin magnetic films: a
review of the classical vector spin model within the mean field
approach}

\author{L. Udvardi\footnote{udvardi@phy.bme.hu}\,\, and L. Szunyogh \\
Department of Theoretical Physics, \\
Budapest University of Technology and Economics, \\
Budafoki \'ut 8, H-1521, Budapest, Hungary \\ 
and \\
Center for Computational Materials Science, \\
Technical University Vienna, \\ 
Gumpendorferstr. 1a, A-1060 Vienna, Austria \\
\ \\
A. Vernes and P. Weinberger \\
Center for Computational Materials Science, \\
Technical University Vienna, \\ 
Gumpendorferstr. 1a, A-1060 Vienna, Austria 
}
\renewcommand{\today}{}
\maketitle
%
%
\begin{abstract}
The ground state and the finite temperature phase diagrams with
respect to magnetic configurations are studied systematically for thin
magnetic films in terms of a classical Heisenberg model including
magnetic dipole-dipole interaction and uniaxial anisotropy.  Simple
relations are derived for the occurrence of the various phase
boundaries between the different regions of the magnetic
orientations. In particular, the range of the first and second order
reorientation phase transitions are determined for bi- and trilayers.
\end{abstract}
%
%
\section{Introduction}
Recent developments of thin film technologies enable the control of
the growth of ultrathin magnetic films deposited on nonmagnetic
substrates. Due to their challenging application as high-storage
magnetic recording media, much attention \cite{G-JMMM86, ASB-PRL90,
PKH-PRL90, PBH-PRB92, A-JMMM94, DFP+PRL94, BH-PRL96, GBO+PRB96,
FPA+PRB97, GCS+PRB97} has been devoted to the novel properties of
these new structures.  From a technological point of view, the study
of the magnetic phase transitions, in particular, of reorientations of
the magnetisation is playing a major role.  As compared to bulk
systems, the presence of surfaces and interfaces leads to an
enhancement of the magneto-crystalline anisotropy due to spin--orbit
coupling.  The magneto-crystalline anisotropy often prefers a
magnetisation perpendicular to the surface, while the magnetic
dipole--dipole interaction and the entropy at finite temperatures
favor an in--plane magnetisation.  Consequently, as observed in many
Fe or Co based ultrathin films \cite{G-JMMM86, ASB-PRL90, PKH-PRL90,
PBH-PRB92, A-JMMM94, DFP+PRL94, BH-PRL96, GBO+PRB96}, a reorientation
from out--of--plane to an in--plane direction of the magnetisation
occurs by increasing both the thickness of the film or the
temperature.  Relativistic first principles calculations using the
local spin density approximation (LSDA) turned out to be sufficiently
accurate to reproduce the critical thickness of the reorientation
\cite{SUW-PRB95,SUW-PRB97,ZSP+PRB98}.

In the case of thin Ni films deposited on a Cu(001) surface, an
opposite behavior was revealed: the magnetisation was found to be
in-plane for less than 7 Ni monolayers, however, it became
perpendicular to the surface beyond this thickness. Below the
switching thickness near 0 K, even an increase of the temperature
induces a similar reversed reorientation \cite{FPA+PRB97,
GCS+PRB97}. The main origin of the above reorientation was attributed
to a strain induced anisotropy of the inner layers preferring a
perpendicular magnetisation \cite{HBW+PRB97,UZW+PRL99}.

Subsequent to the pioneering works of \citet{M:PRB89}, who predicted
the existence of a canted non-collinear ground state for a
semi--infinite ferromagnetic system, and that of \citet{PP:PRL90} who,
by using a renormalization group treatment of a continuum vector field
model, for the first time described the temperature induced (normal)
reorientation phase transition, a considerable amount of theoretical
attempts, mostly by means of different statistical spin models, was
suggested in order to explain the above findings for the thickness and
temperature driven reorientation transitions.  In some attempts,
classical vector spin models were used within the mean field
approximation \cite{TG-JPCM93, HU-JMMM96, HU-PRB97, HU-JMMM99-1,
HU-JMMM99-2, HU-PMB00, JB-SSC98, HLT-PLA99} or in terms of Monte Carlo
simulations \cite{TG-JPCM93, SGL-PRB93, C-PRL95, HMU-JMMM95,
HU-JMMM96, MWD+PRL96}.  A quantum--spin description of reorientation
transitions has also been provided in terms of spin--wave theory
\cite{B-PRB91}, mean field theory \cite{MU-PRB94,MU-PRB95}, many--body
Green function techniques \cite{FJK-EPJB00, FJK+EPJB00, JBB+JAP00},
and by using Schwinger bosonization \cite{TJ-PRB00}.  Although, the
mean field theory is not expected to give a sufficiently accurate
description of low--dimensional systems, it turned out, that it is a
successful tool to study spin reorientation transitions and yields
qualitatively correct predictions \cite{MU-PRB94, MU-PRB95, HU-PRB97,
HU-JMMM99-1, HU-JMMM99-2, HU-PMB00, JB-SSC98, HLT-PLA99}.  It also
should be noted that an itinerant electron Hubbard model revealed the
sensitivity of reorientation transitions with respect to electron
correlation effects \cite{HPN-PRB98}.

For layered systems, the following simple model Hamiltonian can be
used to study reorientation transitions, see e.g. \cite{TG-JPCM93},
\begin{equation}
\begin{aligned} 
 H  = & -{1\over 2}\sum_{(p,i),(q,j)}^{NN} \!\! J \: \vec{s}_{pi} \cdot
\vec{s}_{qj} - \sum_{p,i} \lambda_p(s_{pi}^z)^2 \\ 
& + {1\over 2}\sum_{(p,i) \ne (q,j)} \!\! \omega \left [
{\vec{s}_{pi} \cdot \vec{s}_{qj}\over r_{pi,qj}^3}- 3{(\vec{s}_{pi}
\cdot \vec{r}_{pi,qj})(\vec{s}_{qj} \cdot \vec{r}_{pi,qj}) \over
r_{pi,qj}^5 } \right ] \; ,
\end{aligned}
\label{eq:H}
\end{equation}
where $\vec{s}_{pi}$ ($\mid \! \vec{s}_{pi} \! \mid =1)$ is a
classical vector spin at lattice position $i$ in layer $p$ and
$\vec{r}_{pi,qj}$ is a vector pointing from site $(p,i)$ to site
$(q,j)$ measured in units of the two--dimensional (2D) lattice
constant of the system, $a$.  Although our previous calculations of
the Heisenberg exchange parameters in thin Fe, Co and Ni films on
Cu(001) showed some layer--dependence \cite{SU-PMB98,SU-JMMM99}, in
the first term of Eq.~(\ref{eq:H}) we only consider a uniform nearest
neighbor (NN) coupling parameter $J$ throughout the film.  Similarly,
as we neglect the well--known surface/interface induced enhancement of
the spin--moments, we use a single parameter $\omega=\mu_0\mu^2 / 4\pi
a^3$ (with $\mu_0$ the magnetic permeability and $\mu$ an average
magnitude of the spin--moments), characterizing the magnetic
dipole-dipole interaction strength in the third term of
Eq.~(\ref{eq:H}).  As revealed also by first principles calculations
-- see \citet{WS-CMS00} and references therein -- , the uniaxial
magneto--crystalline anisotropy depends very sensitively on the type
of the surface/interface, the layer--wise resolution of which can vary
from system to system.  Therefore, the corresponding parameters,
$\lambda_p$, in the second term of Eq.~(\ref{eq:H}) remain layer
dependent: the variety of these anisotropy parameters leads to rich
magnetic phase diagrams covering the experimentally detected features
mentioned above.  For example, in a previous study \cite{UKS+JMMM98}
we pointed out that, even in the absence of a fourth order anisotropy
term, for a very asymmetric distribution of $\lambda_p$ with respect
to the layers, the Heisenberg model in Eq.~(\ref{eq:H}) can yield a
canted (non-collinear) ground state.  This feature cannot be obtained
within a phenomenological single domain picture.

As what follows, we first investigate the possible ground states of
the Hamiltonian, E~q.~(\ref{eq:H}). Then we perform a systematic mean
field study of the different kind of temperature induced reorientation
transitions, devoting special attention to the case of bi-- and
trilayers.  Specifically, we define general conditions for the
reversed reorientation.  Most authors in the past focused on proving
the existence of different reorientations and detected only some parts
of the phase diagram, where first and second order phase transitions
occurred.  In here, we describe the full range of uniaxial
anisotropies ($\lambda_p$), for which first or second order
reorientation phase transitions can exist.  Finally, we attempt to
summarize the results and impacts of a mean field approach.

%
%
\section{Ground state}

Confining ourselves to spin--states in which the spins are parallel in
a given layer, but their orientations may differ from layer to layer,
i.e.
\begin{equation} \label{eq:sp}
\vec{s}_{pi} = \vec{s}_p = \left( \sin(\theta_p) \cos(\phi_p),
\sin(\theta_p) \sin(\phi_p),\cos(\theta_p) \right) \quad ,
\end{equation}
where $\theta_p$ and $\phi_p$ are the usual azimuthal and polar angles
with the $z$ axis normal to the planes, the energy of $N$ layers per
2D unit cell can be written as
\begin{equation}
\begin{aligned} 
\, & E_N ( \theta_1, \theta_2, \dots, \theta_N; \phi_1, \phi_2, \dots,
\phi_N) = 
 -{1\over 2}\sum_{p,q=1}^N (Jn_{pq} - A_{pq}\omega) \cos{(\theta_p)}
\cos{(\theta_q)} \\ 
& -{1\over 2}\sum_{p,q=1}^N (Jn_{pq} +{1\over 2} A_{pq}\omega)  
\sin{(\theta_p)} \sin{(\theta_q)} \cos(\phi_p -\phi_q) 
-\sum_{p=1}^N \lambda_p\cos^2{(\theta_p)} \quad ,
\end{aligned}
\label{eq:E1}
\end{equation}
with $n_{pq}$ being the number of nearest neighbors in layer $q$ of a
site in layer $p$, and $A_{pq}$ the magnetic dipole--dipole coupling
constants, see Appendix in \cite{SUW-PRB95},
\begin{equation} \label{eq:Apq}
{\sum_j}^\prime \frac{1}{r_{p0,qj} ^3} \left ( I - 
3{\vec{r}_{p0,qj} \otimes \vec{r}_{p0,qj} \over r_{p0,qj}^2 }
 \right ) = A_{pq} \times \left( \begin{array}{ccc}
                            -{1\over 2} & 0           &  \; 0  \\
                              0         & -{1\over 2} &  \; 0  \\
                              0         &  0          &  \; 1  \\
                           \end{array} \right )   \quad ,
\end{equation}
which is valid for square and hexagonal 2D lattices, such as the (100)
and (111) surfaces of cubic systems. ($I$ the three dimensional unit
matrix, while $\otimes$ stands for the tensorial product of two
vectors.)  For three--dimensional (3D) translational invariant
underlying parent lattices \cite{W-PMB97}, $A_{pq}$ depends only on
$|p - q|$, i.e. the distance between layers $p$ and $q$.  In table
\ref{tbl:mad} we summarize these constants for the first few layers of
the most common cubic structures.  As the magnetic dipole--dipole
interaction is clearly dominated by the positive first (and second)
layer couplings, for ferromagnetic systems ($J > 0$), a minimum of the
energy in Eq.~(\ref{eq:E1}) corresponds to the case when all polar
angles, $\phi_p$, are identical.  Therefore, the $\phi$--dependence in
Eq.~(\ref{eq:E1}) disappears and the expression,
\begin{equation}
\begin{aligned} 
E_N(\theta_1, \theta_2, \dots, \theta_N) = & 
 -{1\over 2}\sum_{p,q=1}^N (Jn_{pq} - A_{pq}\omega)
\cos{(\theta_p)} \cos{(\theta_q)} \\
& -{1\over 2}\sum_{p,q=1}^N (Jn_{pq} +{1\over 2} A_{pq}\omega) 
\sin{(\theta_p)} \sin{(\theta_q)} 
-\sum_{p=1}^N \lambda_p\cos^2{(\theta_p)} \quad ,
\end{aligned}
\label{eq:E2}
\end{equation}
has to be minimized with respect to the $\theta_p$.  The corresponding
Euler--Lagrange equations are then
\begin{equation}
\begin{aligned}
{\partial E_N(\theta_1, \theta_2, \dots, \theta_N) \over
\partial\theta_p} = & 
\ \ \,\sum_{q=1}^N (Jn_{pq} -A_{pq}\omega)\sin{(\theta_p)}\cos{(\theta_q)}
\\ 
& -\sum_{q=1}^N (Jn_{pq} +{1\over 2}
A_{pq}\omega)\cos{(\theta_p)}\sin{(\theta_q)} \\
& + 2\lambda_p\sin{(\theta_p)}\cos{(\theta_p)} = 0 \; .
\end{aligned}
\label{eq:EL}
\end{equation}
Obviously, a uniform in--plane ($\{ \theta_p = \pi / 2 \}$) and a
normal--to--plane ($\{ \theta_p = 0 \}$) orientation satisfy
Eq.~(\ref{eq:EL}). The energies of these two particular spin--states
coincide, if
\begin{equation} \label{eq:epn}
 \sum_{p=1}^{N} {\lambda_p\over\omega} = {3\over 4}
\sum_{p,q=1}^{N} A_{pq}\; ,
\end{equation}
which defines an $(N\!-\!1)$--dimensional hyper--plane in the
$N$--dimensional space of parameters $\{ {\lambda_p / \omega}\}$.  If
the magnetisation changes continuously across this plane, in its
vicinity there should exist solutions with canted magnetisation.
Moreover, the saddle points of the energy functional in
Eq.~(\ref{eq:E2}),
\begin{equation}
 det \left | {\partial E_N\over \partial\theta_p \partial\theta_q}
\right |_{ \{ \theta_p=0,{\pi\over 2} \} } = 0 \;,
\end{equation}
define the boundaries of the canted zone, 
\begin{equation} \label{eq:b1}
 det\left | \left[ \sum_{r=1}^N \left({J \over \omega} n_{pr}+{1\over
2} A_{pr}\right) - 2 {{\lambda_p}\over{\omega}} \right] \delta_{pq} -
\left( {J \over \omega} n_{pq} - A_{pq} \right) \right | = 0 \; ,
\end{equation}
and 
\begin{equation} \label{eq:b2}
 det\left | \left[ \sum_{r=1}^N\left( {J \over \omega} n_{pr}- A_{pr}
\right) + 2 {{\lambda_p}\over{\omega}} \right] \delta_{pq} - \left({J
\over \omega} n_{pq} +{1\over 2} A_{pq}\right) \right| = 0 \; ,
\end{equation}
for the uniform in--plane and normal--to--plane magnetisations,
respectively.  For a bilayer we derived explicit expressions of
Eqs.~(\ref{eq:b1}) and (\ref{eq:b2}), see \citet{UKS+JMMM98}.

In order to study the canted region, instead of solving the
Euler--Lagrange equations, Eq.~(\ref{eq:EL}), directly, we fixed a
configuration $\theta_1^\star, \theta_2^\star\dots \theta_N^\star$ and
determined $\lambda_1,\lambda_2 \dots \lambda_N$ by demanding that
Eq. (\ref{eq:EL}) must be satisfied, i.e.
\begin{equation} \label{eq:lambda}
\lambda_p = {1\over 2} \sum_{q=1}^N (Jn_{pq}+{1\over 2} A_{pq}\omega)
             {\sin(\theta_q^\star)\over \sin(\theta_p^\star)} -
             {1\over 2} \sum_{q=1}^N (Jn_{pq}-A_{pq}\omega)
             {\cos(\theta_q^\star)\over \cos(\theta_p^\star)} \; .
\end{equation}
Substituting these parameters into Eq.~(\ref{eq:E2}), one easily can
express the difference of the energies between the corresponding
configurations as,
\begin{eqnarray} \label{eq:Ediff1}
&& E_N(\theta_1=\theta_2=\dots = \theta_N=0) - 
E_N(\theta_1^{\star},\theta_2^{\star},\dots \theta_N^\star) = \\
\nonumber
&& {1\over 2}\sum_{pq}^{N}
      \left ( n_{pq}J - A_{pq}\omega \right )
      \left[ {\cos{(\theta_p^\star)}\over \cos{(\theta_q^\star)}} +
             {\cos{(\theta_q^\star)}\over \cos{(\theta_p^\star)}} - 2
\right ] ,
\end{eqnarray}
and
\begin{eqnarray} \label{eq:Ediff2}
&& E_N(\theta_1=\theta_2=\dots = \theta_N={\pi\over 2}) - 
E_N(\theta_1^{\star},\theta_2^{\star},\dots \theta_N^\star) = \\
\nonumber
&& {1\over 2}\sum_{p,q}^{N}
      \left ( n_{pq}J + {1\over 2} A_{pq}\omega \right )
      \left[ {\sin{(\theta_p^*)}\over \sin{(\theta_q^*)}} +
             {\sin{(\theta_q^*)}\over \sin{(\theta_p^*)}} - 2
\right] \; .
\end{eqnarray}
Note that the position of the minimum $\{ \theta_p^{\star}\}$ as well
as the minimum energy $E_N(\{ \theta_p^{\star} \})$ are functions of
the parameters $J$, $\omega$, and $\{ \lambda_p \}$.  Obviously,
whenever the parameters fall into the region between the two
hyper--planes defined by Eqs.~(\ref{eq:b1}) and (\ref{eq:b2}), the
energy of non--collinear states is always below or equal to the energy
of the collinear in-plane or normal--to--plane solutions.

We showed that for a bilayer, $\lambda_1 =\lambda_2 = 3
(A_{11}+A_{12}) \omega / 4$ implies a collinear ground state spin
configuration \cite{UKS+JMMM98}.  This state is, however, continuously
degenerate, i.e. the energy is independent of the orientation of the
magnetisation.  Such a critical point in the phase diagram also exists
for multilayers ($N \ge 3$).  Namely, from Eqs.~(\ref{eq:Ediff1}) and
(\ref{eq:Ediff2}) it follows that for $\theta_1^\star = \theta_2^\star
\dots = \theta_N^\star = \theta^\star$, $E_N(\{ \theta_p^\star \})$ is
independent of $\theta^\star$.  In terms of Eq.~(\ref{eq:lambda}), the
corresponding point in the parameter space $\{ \lambda_p / \omega \}$
is given by
\begin{equation} \label{eq:crit}
 {\lambda_p\over\omega} = {3\over 4}\sum_{q=1}^N A_{pq} \; .
\end{equation}
Evidently, the hyper--planes given by Eq.~(\ref{eq:b1}) and
Eq.~(\ref{eq:b2}) touch the hyper--plane, separating the in--plane and
normal--to--plane regions, Eq.~(\ref{eq:epn}), at the point defined by
Eq.~(\ref{eq:crit}).  It is worthwhile to mention that this is the
only point where canted collinear solutions can exist.  This critical
point was also found by \citet{HU-JMMM96} for a monolayer, but they
did not prove its existence for multilayers.
%
%
\section{Finite temperature}
Introducing the following coupling constants 
\begin{equation} \label{eq:cpxz}
c^x_{pq} = n_{pq}J + {1\over 2}\omega A_{pq} \quad {\rm and} \quad
c^z_{pq} = n_{pq}J - \omega A_{pq} \; ,
\end{equation}
the molecular field corresponding to the Hamiltonian Eq.~(\ref{eq:H}) 
at layer $p$ can be written as
\begin{equation} 
\begin{aligned}
 H^p(\theta,\phi) = & - \sum_{q=1}^N  c^x_{pq} 
\left[ m^x_p \sin{(\theta)}\cos{(\phi)} +
 m^y_p \sin{(\theta)}\sin{(\phi)} \right] \\
 & - \sum_{q=1}^N c^z_{pq} m^z_q \cos{(\theta)} - 
   \lambda_p \cos^2{(\theta)}
\; ,
\end{aligned}
\label{eq:mf}
\end{equation}
where $m^\alpha_p = \langle s^\alpha_p \rangle$ ($\alpha=x,y,z$).
Similar to the ground state (see Section II.), due to the in--plane
rotational symmetry of the above effective Hamiltonian, the in--plane
projections of all the average magnetic moments $\vec{m}_p$ are
aligned. Therefore, by choosing an appropriate coordinate system,
$m^y_p$ can be taken to be zero in Eq.~(\ref{eq:mf}).  The partition
function is then defined by
\begin{equation} \label{eq:part}
 Z = \prod_{p=1}^N Z_p \; ,
\end{equation}
\begin{equation}
 Z_p = 2\pi\int\limits_{-{\pi \over 2}}^{\pi \over 2}
 \exp\left\{ \beta \left[ b_p^z\cos(\theta) +
\lambda_p \cos^2(\theta) \right] \right\} 
J_0(-i\beta b_p^x \sin(\theta)) \sin(\theta)d\theta \; ,
\end{equation}
where 
\begin{equation} \label{eq:bpxz}
b^{x(z)}_p = \sum_{q=1}^N c^{x(z)}_{pq}m^{x(z)}_q \; ,
\end{equation}
$\beta = 1/(k_B T)$, $k_B$ the Boltzmann--constant and $T$ the
temperature.  The minimization of the free--energy with respect to the
average magnetisations leads to the following set of nonlinear
equations
\begin{equation} 
\label{eq:mfx}
m_p^x = {2i\pi\over Z_p}\int\limits_{-{\pi \over 2}}^{\pi \over 2}
\sin(\theta) \exp\left\{ \beta \left[ b_p^z\cos(\theta) + \lambda_p
\cos^2(\theta) \right] \right\} J_1(-i\beta b_p^x \sin(\theta))
\sin(\theta)d\theta
\end{equation}
and
\begin{equation} \label{eq:mfz}
 m_p^z = {2\pi\over Z_p}\int\limits_{-{\pi \over 2}}^{\pi \over 2}
 \cos(\theta) \exp\left\{ \beta \left[ b_p^z\cos(\theta) + \lambda_p
 \cos^2(\theta) \right] \right\} J_0(-i\beta b_p^x \sin(\theta))
 \sin(\theta)d\theta \; .
\end{equation}
In Eqs.~(\ref{eq:part}), (\ref{eq:mfx}) and (\ref{eq:mfz}), $J_0$ and
$J_1$ denote Bessel functions of zero and first order, respectively
\cite{ASbook-70}.

By using a high temperatures expansion, Eqs.~(\ref{eq:mfx}) and
(\ref{eq:mfz}) become decoupled (see Appendix).  Consequently, the
magnetisation can go to zero either via an in--plane or via a
normal--to--plane direction at temperatures $T_x$ and $T_z$,
respectively, and the higher one of them can be associated with the
Curie temperature $T_C$.  Clearly, an out--of--plane to in--plane
reorientation phase transition can occur only when the ground state
magnetisation is out--of--plane and $T_z < T_x = T_C$. In the case of
a reversed reorientation transition, the ground state magnetisation
has to be in--plane (or canted) and $T_x < T_z = T_C$.

Expanding $T_x$ and $T_z$ up to first order of the anisotropy
parameters $\lambda_p$, leads to the following expressions (see
Appendix)
\begin{equation}
\begin{aligned}
T_x & = {1\over {3 k_B}}\left[ n_{11}J + {1\over 2} A_{11}\omega +
2\left( n_{12}J + {1\over 2} A_{12}\omega \right) \cos\left (
{\pi\over N+1}\right ) \right] \\ & - {4\over {15(N+1)}} \sum_{p=1}^N
\lambda_p \sin^2\left ({p\pi\over N+1}\right ) \; ,
\end{aligned}
\label{eq:tcx}
\end{equation}
and 
\begin{equation}
\begin{aligned}
T_z & = {1\over {3 k_B}}\left[ n_{11}J - A_{11}\omega + 2\left(
n_{12}J - A_{12}\omega \right) \cos\left ( {\pi\over N+1}\right )
\right] \\ & + {8 \over {15(N+1)}} \sum_{p=1}^N \lambda_p \sin^2\left
({p\pi\over N+1}\right ) \;.
\end{aligned}
\label{eq:tcz}
\end{equation}
The above expressions imply that, if the anisotropy parameters
$\lambda_p$ are small, $T_x$ is larger than $T_z$.  As the anisotropy
parameters are increasing, the difference between $T_z$ and $T_x$
decreases.  The two temperatures coincide, if the following condition
is fulfilled,
\begin{equation} \label{eq:Txz}
 \sum_{p=1}^N \left( {\lambda_p\over \omega} \right) 
  \sin^2 \left( {p\pi\over N+1}\right) = 
 {{5(N+1)} \over 8} \left[  A_{11} + 2A_{12} \cos\left ( {\pi\over
N+1} \right ) \right]  \; .
\end{equation}
Above the hyper--plane determined by Eq.~(\ref{eq:Txz}), i.e. for $T_x
< T_z$, the uniaxial anisotropy is large enough to keep the
magnetisation normal to the surface as long as the temperature reaches
$T_C$.

First principles calculations on (Fe,Co,Ni)/Cu(001) overlayers
revealed \cite{USW-PRB96, SUB+PRB97, SU-PMB98, SU-JMMM99, UZW+PRL99},
that the uniaxial magnetic anisotropy energy and the magnetic
dipole--dipole interaction are two to three orders of magnitude
smaller than the exchange coupling.  Thus, for physically relevant
parameters, the boundaries of the canted ground state fixed by
Eqs.~(\ref{eq:b1}) and (\ref{eq:b2}) are close to the hyper--plane
defined by Eq.~(\ref{eq:epn}). Apart form this tiny range of canted
ground states, temperature induced out--of--plane to in--plane
reorientation can occur in the parameter space $\{ \lambda_p / \omega
\}$ between the two hyper--planes given by Eqs.~(\ref{eq:epn}) and
(\ref{eq:Txz}). It is worthwhile to mention that the positions of
these hyper--planes are determined only by the magnetic dipole--dipole
constants $A_{pq}$.

\begin{figure}[htbp] \centering
\includegraphics[width=0.9\columnwidth,clip]{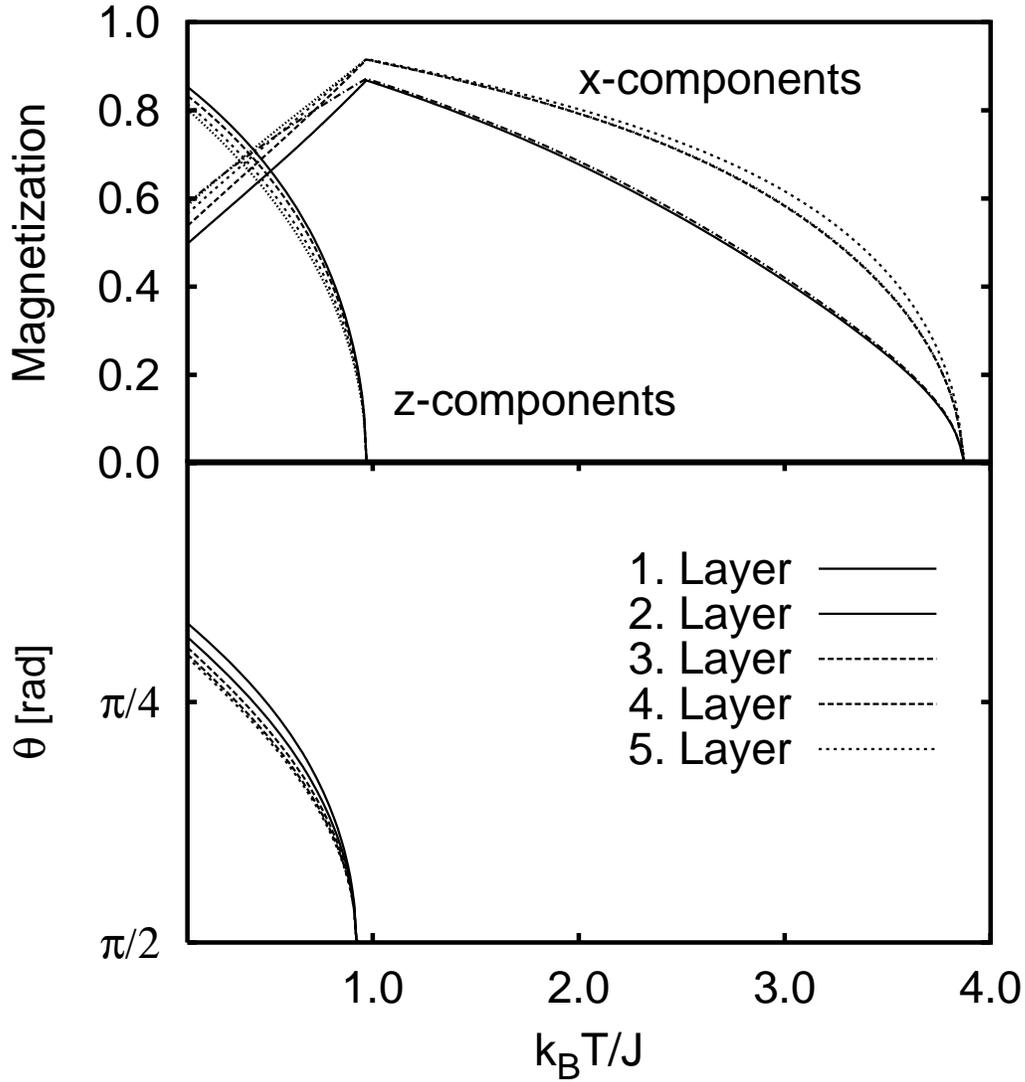}
\caption{ \label{fig1}
Out--of--plane to in--plane reorientation transition in a 5-layer
system ($\lambda_1/J=0.26$, $\lambda_{2,\dots,5} = 0$, $\omega/J =
0.0056$).  The $z$-- and the $x$--components of the average
magnetisation as well as the angles of the magnetisation with
respect to the normal of the surface are shown in the upper and the
lower panels, respectively. 
}
\end{figure}

An example for an out--of--plane to in--plane reorientation transition
in a 5--layer thick film is shown in figure \ref{fig1}.  Neglecting
the fourth order anisotropy terms, the parameters of the system have
been chosen identical to those characteristic to a Co$_5$/Au(111)
overlayer \cite{UKS+JMMM98}.  Due to the highly asymmetric
distribution of the $\lambda_p$ with respect to the layers, the system
has a non--collinear canted ground state. As the temperature
increases, the magnetisation in each layer turns into the plane of the
film.  The system keeps its non--collinear configuration up to the
reorientation transition temperature ($\sim 0.9 J/k_B$), above which
it is uniformly magnetized in--plane up to the Curie temperature
($\sim 3.8 J/k_B$).

\begin{figure}[htbp] \centering
\includegraphics[width=0.8\columnwidth,clip]{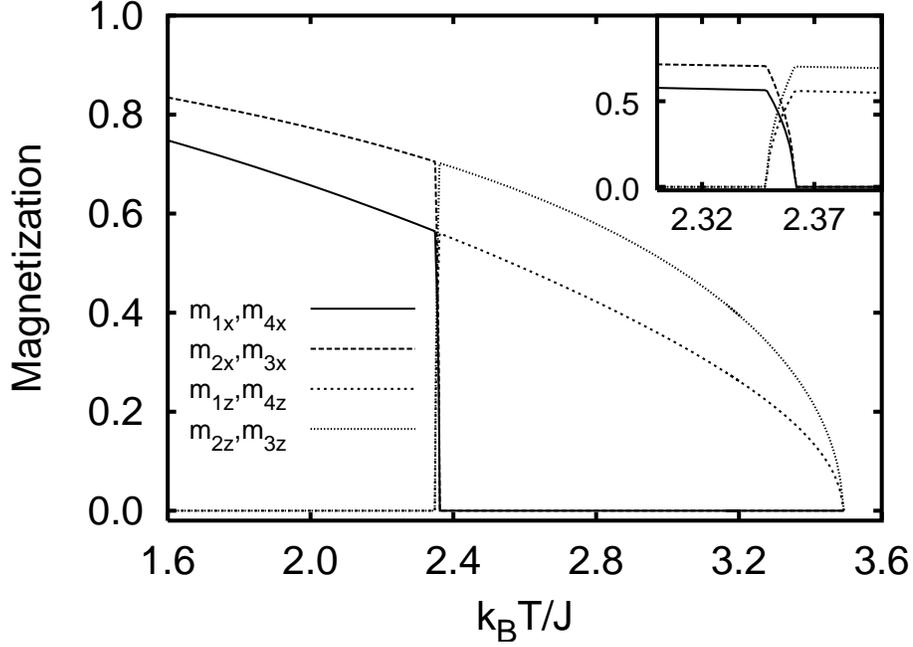}
\caption{ \label{fig2} Normal--to--plane and in--plane components of
the layer resolved magnetisation for a film of 4 atomic layers
exhibiting a reversed reorientation transition. The parameters,
representative to Ni were taken from \protect{\citet{HU-PRB97}}:
$J=1$, $\lambda_1/J=-3.5\times 10^{-3}$, $\lambda_i/J=1.5\times
10^{-3} \; (i>1) $, $\omega/J=5\times 10^{-5}$.  The inset shows the
vicinity of the reorientation transition on an enlarged scale. }
\end{figure}

\begin{figure}[htbp] \centering
\includegraphics[width=0.6\columnwidth,clip]{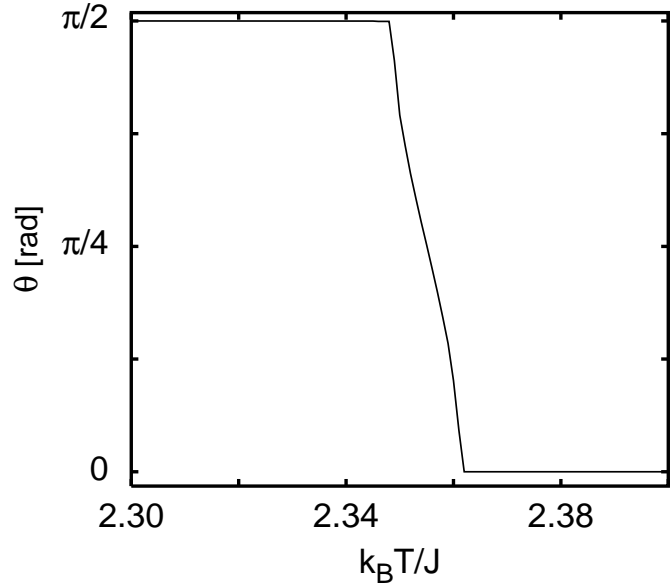}
\caption{ \label{fig3}
Variation of the angle of the average magnetisation for the system in
Fig. \protect{\ref{fig2}}. }
\end{figure}

The temperature induced reversed reorientation transition, found
experimentally in Ni$_n$ / Cu(001) films for $n < 7$, has successfully
been described by \citet{HU-PRB97}, who used a perturbative mean field
approach to the model given in Eq.~(\ref{eq:H}).  Using the same
parameters, we solved the mean field equations (\ref{eq:mfx}) and
(\ref{eq:mfz}) and reproduced the reversed reorientation transition
without any perturbative treatment. The results for a 4--layer film
are shown in figures \ref{fig2} and \ref{fig3}.  Although the
distribution of the anisotropy parameters is asymmetric, the
calculation resulted in identical magnetisations in the first and
fourth layer as well as in the second and third layer. Moreover, the
angles of the magnetisations in the different layers are almost
identical: in the whole temperature range the largest deviation is
smaller than $6\times 10^{-4}$ rad.

For a bilayer ($N=2$) the hyper-planes (\ref{eq:epn}) and
(\ref{eq:Txz}) reduce to the lines
\begin{equation}
 {\lambda_1\over \omega} + {\lambda_2\over\omega} = 
        {3\over 2} \left ( A_{11} + A_{12} \right ) \quad
{\rm and} \quad
 {\lambda_1\over \omega} + {\lambda_2\over\omega} = 
        {5\over 2} \left ( A_{11} + A_{12} \right ) \; , 
\end{equation}
respectively. Apparently, the two lines do not intersect.  As a
consequence, a reversed reorientation can occur only if the number of
layers in the film exceeds two.  The same conclusion has been drawn by
\citet{HU-PRB97} using a perturbative treatment of the anisotropy
parameters.  Nevertheless, it is interesting to note, that the region
in the parameter space $\{ \lambda_p / \omega \}$ of canted ground
states, bounded by the lines defined by Eqs.~(\ref{eq:b1}) and
(\ref{eq:b2}), always overlaps the region, where the magnetisation
goes to zero via in--plane orientation.  Thus, in this overlapping
region an out--of--plane (canted) to in--plane, i.e. reversed
reorientation transition can indeed occur.  The corresponding
parameters, $\omega$ and $\{ \lambda_p \}$, are, however, most likely
beyond the physically relevant regime.

\begin{figure}[t] \centering
\includegraphics[width=0.95\columnwidth,clip,bb=5 0 404 281]{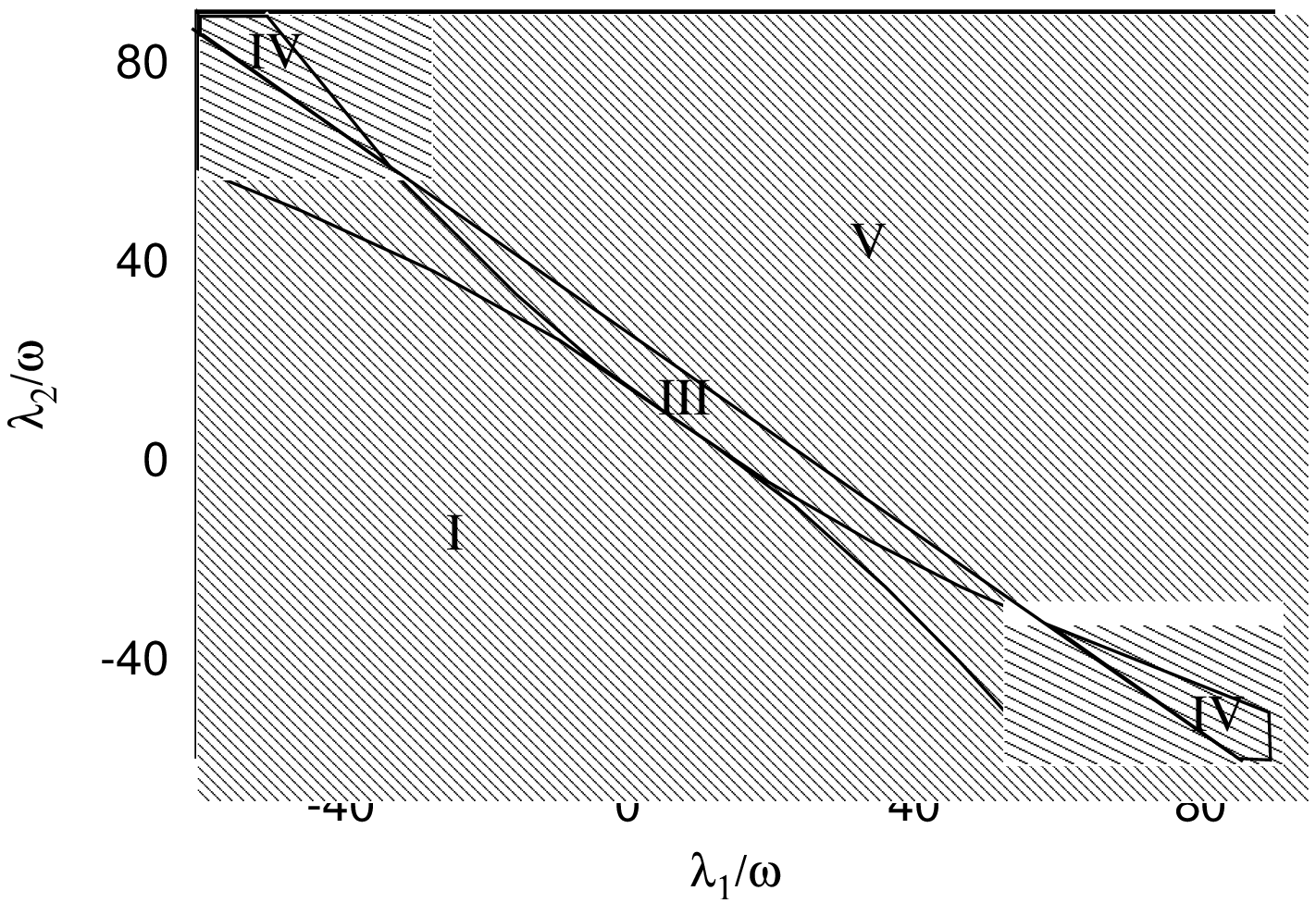}
\vskip 0.3cm
\caption{ \label{fig4}
Phase diagram of the magnetic ground states and the reorientation
phase transitions for a bilayer. The magnetic dipole--dipole coupling
strength $\omega=0.01J$.  I: in--plane magnetisation up to
$T_C$; II: canted ground state with reorientation transition to
in--plane orientation; III: normal--to--plane ground state with
reorientation transition to in--plane orientation; IV: canted ground
state with reversed reorientation transition; V: normal--to--plane
magnetisation up to $T_C$. }
\end{figure}

For a bilayer, in figure \ref{fig4} the different regions of phase
transitions in the respective parameter space are shown.  In regions I
and V there is no temperature driven reorientation transition and the
magnetisation remains in--plane and normal--to--plane, respectively,
until $T_C$ is reached. In the regions II and III, the magnetisation
turns into the plane from a canted or a normal--to--plane ground
state, respectively. As discussed above, in region IV, a reversed
reorientation can occur from a canted ground state to a
normal--to--plane direction.

The order of the reorientation transition at finite temperatures has
been studied in the literature by mean field and Monte Carlo methods.
Most authors concluded \cite{HMU-JMMM95, HU-JMMM96, MWD+PRL96} that
the reorientation transition in a monolayer is of first order. For a
bilayer, within the mean field approach, a relatively small range in
the vicinity of $\lambda_1=\lambda_2$ was found, where the system
underwent a first order reorientation transition \cite{HU-JMMM96}.  In
what follows, we establish a simple, general criterion for the order
of the reorientation transition. Suppose that the ground state
magnetisation is in--plane and its normal--to--plane component appears
at the temperature $T_{rz}$.  Since near $T_{rz}$ the $z$--component
of the magnetisation is small, the exponential function in
Eq.~(\ref{eq:mfz}) can be expanded up to first order in $m^z_p$,
leading to the homogeneous linear equations
\begin{equation} \label{eq:rz}
 \sum_{q=1}^N C^z_{pq} m^z_{q} = 0 \quad ,
\end{equation}
where
\begin{equation} 
 C^z_{pq} \equiv \delta_{pq} - \beta_{rz} c^{z}_{pq}{\mathcal Z}_p \quad ,
\end{equation}
\begin{equation} 
 {\mathcal Z}_p = {2\pi\over Z_p}\int \cos^2(\theta) 
  \exp\left [ \beta_{rz} \lambda_p \cos(\theta)^2\right ]
  J_0(-i\beta_{rz} b_p^x \sin(\theta)) \sin(\theta)d\theta \quad ,
\end{equation}
and $\beta_{rz} \equiv 1 / (k_B T_{rz})$.  Note, that the $C^z_{pq}$
depend only on the in-plane component of the magnetisations $m^x_{q}$,
which has to satisfy Eq.~(\ref{eq:mfx}) for the case of $m^z_q = 0$.
Eq.~(\ref{eq:rz}) has a non-trivial solution only, if the determinant
of the matrix $C^z = \left\{ C^z_{pq} \right\}$ is zero.  Evidently,
this is the condition which determines $T_{rz}$.  Similarly, one can
easily find the corresponding equation for $T_{rx}$, where the
in--plane component of the magnetisation appears in a
normal--to--plane spin configuration,
\begin{equation} \label{eq:rx}
 \sum_{q=1}^N C^x_{pq} m^x_{q} = 0 \quad , 
\end{equation}
with
\begin{equation} 
 C^x_{pq} = \delta_{pq} - \beta_{rx} c^{x}_{pq} {\mathcal X}_p \quad , 
\end{equation}
\begin{equation} 
 {\mathcal X}_p = {\pi\over Z_p}\int \sin^3(\theta) 
  \exp\left [ \beta_{rx} \lambda_p \cos^2(\theta) \right ]
  \left[ J_0( -i\beta_{rx} b_p^x \sin(\theta)) -
   J_2( -i\beta_{rx} b_p^x \sin(\theta)) \right]
  d\theta \; ,
\end{equation}
and $\beta_{rx} \equiv 1 / (k_B T_{rx})$.  It is easy to show, that
Eqs.~(\ref{eq:rz}) and (\ref{eq:rx}) directly follow from a stability
analysis of the mean field free energy in the vicinity, where the
corresponding components of the magnetisations vanish. 

\begin{figure}[htbp] \centering
\includegraphics[width=0.9\columnwidth,clip]{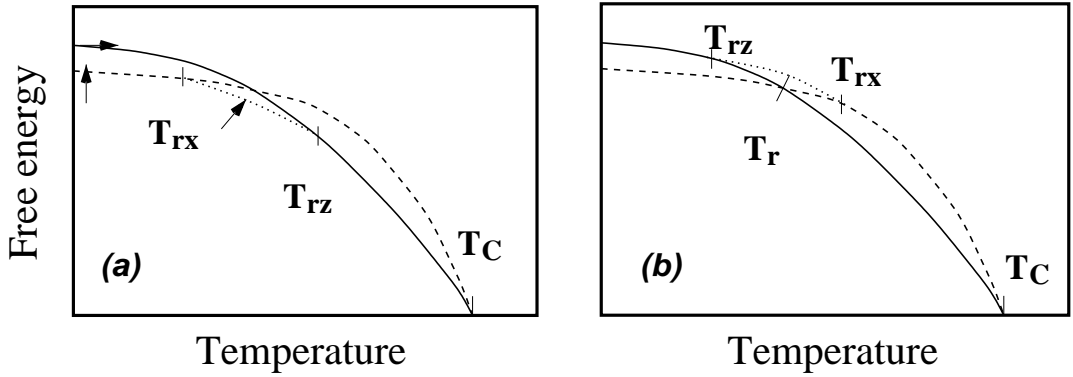}
\caption{ \label{fig5}
Schematic picture of the free--energy in the case of a second order
({\it a}) and a first order ({\it b}) reorientation phase
transition. As indicated by arrows, the solid, dashed
and dotted lines refer to the in--plane, normal--to--plane and canted
mean field solutions, respectively. }
\end{figure}

The mean field equations, Eqs.~(\ref{eq:mfx}) and (\ref{eq:mfz}),
always have an in--plane and a normal--to--plane solution with
magnetisations $m^q_x \neq 0$, $m^q_z = 0$ and $m^q_z \neq 0$, $m^q_x
= 0$, respectively. Between $T_{rx}$ and $T_{rz}$, a canted solution
can exist with $m^q_x \neq 0$ and $m^q_z \neq 0$.  Among the above
three phases, the physical one belongs to that, which has the lowest
free--energy.  In figure \ref{fig5}$a$, the free--energy of a system
possessing a second order normal--to--plane to in--plane reorientation
transition is schematically shown.  The ground state magnetisation is
perpendicular to the surface of the substrate.  At $T_{rx}$ an
in--plane component appears in the magnetisation.  The
normal--to--plane component of the magnetisation vanishes at the
temperature $T_{rz} (> T_{rx})$.  A similar picture for a first order
transition is shown in figure \ref{fig5}b. Obviously, one can conclude
that, if $T_{rx} < T_{rz}$, a second order normal--to--plane to
in--plane reorientation phase transition occurs, whereas, if $T_{rz} <
T_{rx}$, the reorientation transition is of first order.  In the case
of a reversed reorientation, the relation between $T_{rx}$ and
$T_{rz}$ is just the opposite as before: a second order transition
occurs, if $T_{rz} < T_{rx}$, while for $T_{rx} < T_{rz}$ the
transition is of first order. At the boundary of the regions, where
second order and first order phase transitions occur, the two
temperatures, $T_{rx}$ and $T_{rz}$, must evidently coincide.

\begin{figure}[htbp] \centering
\includegraphics[width=0.9\columnwidth,clip]{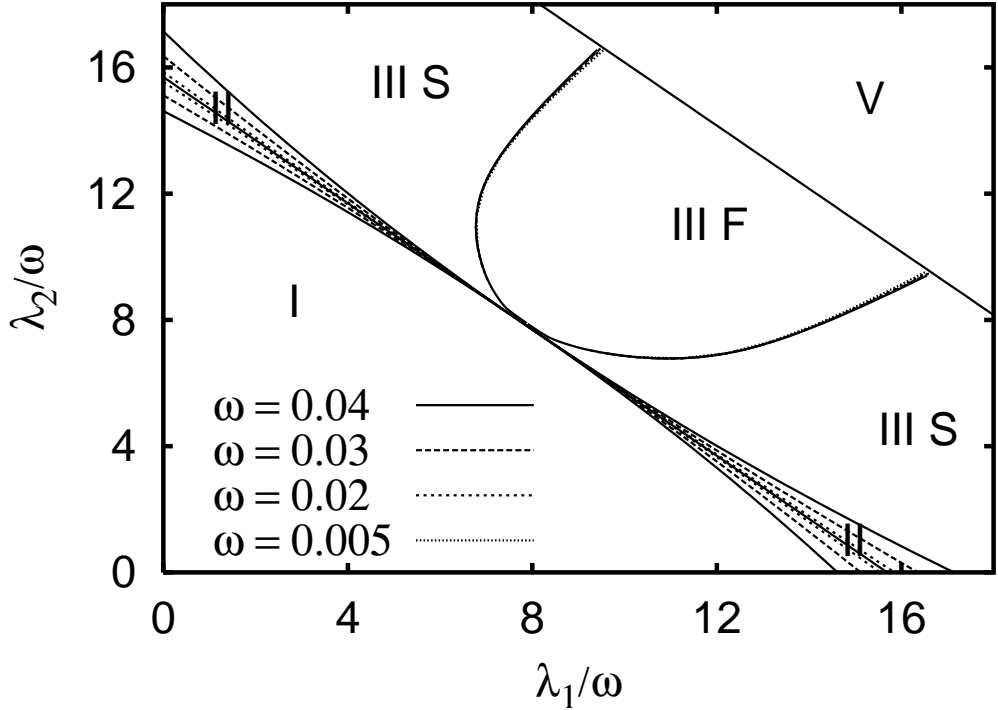}
\caption{ \label{fig6} 
Phase diagram of reorientation transitions for a bilayer ($N=2$).
Labels I, II, III and V denote the same regions as in figure
\protect{\ref{fig4}}; regime III, however, is partitioned into regions
referring to first (III F) and second (III S) order normal
reorientation phase transitions.  The corresponding boundary lines are
shown for different values of the magnetic dipole--dipole interaction
strength $\omega$, measured in units of $J$.  }
\end{figure}

In figure \ref{fig6}, the region of first order reorientations (III~F)
and that of second order reorientations (III~S) are shown in the phase
diagram for a bilayer.  Note, that figure \ref{fig6} in fact
represents figure \ref{fig4} on an enlarged scale for the parameters
$0 < \lambda_{1,2}/\omega < 18$.  This picture is consistent with the
observation of \citet{HU-JMMM96} for the range of the first order
reorientation phase transitions, as they performed investigations very
close to the critical point only. In that case, by keeping
$\lambda_1+\lambda_2$ fixed, figure \ref{fig6} implies a very narrow
range for the first order transitions.

\begin{figure}[htbp] \centering
\includegraphics[width=0.9\columnwidth,clip]{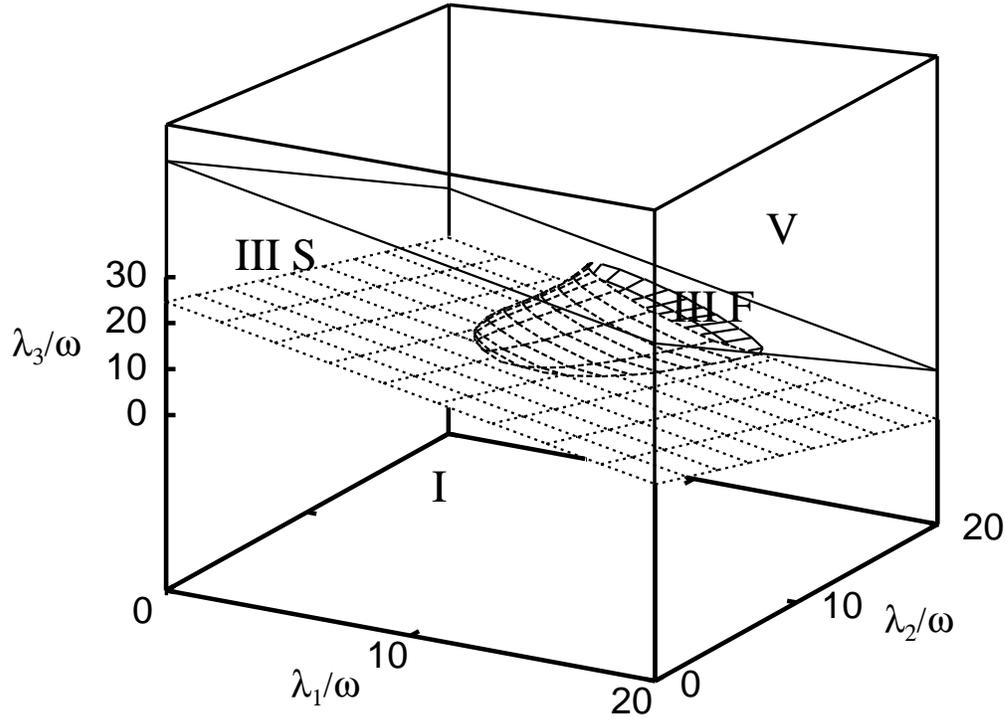}
\caption{ \label{fig7}
Mean--field phase diagram of reorientation transitions for a trilayer
($N=3$).  I: in--plane magnetisation up to $T_C$, III S: second order
normal--to--plane to in--plane reorientation, III F: first order
normal--to--plane to in--plane reorientation, V: normal--to--plane
magnetisation up to $T_C$. }
\end{figure}

The phase diagram of the trilayer case ($N=3$) is shown in figure
\ref{fig7}.  Apparently, the same regimes exist as in the case of a
bilayer. The region of first order reorientation transition forms now
a 'sack', touching the plane defined by Eq.~(\ref{eq:epn}) at the
critical point given by Eq.~(\ref{eq:crit}).  The sack is covered by
the plane separating the area where normal--to--plane to in--plane
reorientation occur and the area, where the magnetisation remains
normal--to--plane up to the Curie temperature, see Eq.~(\ref{eq:Txz}).
The regime of reversed reorientation transitions, part of the region
of second order transitions, is, however, out of the segment of the
parameter space depicted in figure \ref{fig7}.

Numerical calculations using different magnetic dipole-dipole coupling
strengths $\omega$ (see, in particular, figure \ref{fig6}) yields
practically the same boundaries in the $\{ \lambda_p / \omega \}$
parameter space of the phase diagrams for both the bilayer and
trilayer cases.  The only exception is the region of the canted ground
states, which rapidly opens up with increasing $\omega$.  The
established {\em universality} of the phase boundaries nicely confirms
that the reorientation phase transitions, as long as $\omega$ gets
comparable to $J$, are a consequence of the competition between the
uniaxial anisotropy and the magnetic dipole--dipole interaction.
%
%
\section{Conclusions}

In the present paper we provided a full account of the ground states
and of the finite temperature behavior of a ferromagnetic film of
finite number of layers, as described by the classical vector spin
Hamiltonian, Eq.~(\ref{eq:H}), including exchange coupling
interaction, uniaxial magneto--crystalline anisotropies and magnetic
dipole--dipole interaction.  We derived explicit expressions for the
boundaries of the regions related to normal--to--plane, canted and
in--plane ground states in the corresponding parameter space. We
concluded that within the model, defined by Eq.~(\ref{eq:H}), canted
ground states are ultimately connected to non--collinear
spin--configurations. In addition -- so far established for monolayers
only \cite{HU-JMMM96} -- for any thickness of the film we proved the
existence of a critical point, where the ground state energy of the
system is independent from a uniform orientation of the magnetisation.

We also investigated intensively the finite temperature behavior of
the system in terms of a mean field theory. By using a high
temperature expansion technique, we showed that the Curie temperature
of a ferromagnetic film can be calculated by solving an eigenvalue
problem, which, for the case of a bulk system and by neglecting
anisotropy effects, leads to the well--known expression of $T_C$.  The
main part of the present study has been devoted to the reorientation
phase transitions, which play a central role for applications of thin
film and multilayer systems as high--storage magnetic recording
devices. Both the normal--to--plane to in--plane and the in--plane to
normal--to--plane (reversed) temperature induced reorientation
transitions have been discussed and the corresponding regions in the
parameter space have been explicitly determined.  In accordance to
previous studies \cite{HU-PRB97}, we showed that, for physically
relevant parameters, reversed reorientation can occur only for films
containing three or more atomic layers.  By investigating the order of
reorientation phase transitions, we found well--defined conditions for
the first and the second order phase transitions and presented the
corresponding regions for bi-- and trilayers in the respective
parameter spaces.

In conclusion, we have shown that a mean field treatment of a
classical vector spin model recovers most of the important phenomena
observed in magnetic thin film measurements at finite
temperatures. Without any doubt, due to the lack of mean field
theories for low--dimensional systems, some of them have to be refined
by using more sophisticated methods of statistical physics (see
Introduction).  In particular, for very thin films (monolayers), the
mean field theory predicts a $T_C$ much higher than the random phase
approximation (RPA).  However, by rescaling the temperature, the
orientations of the magnetisation become fairly similar in both
approaches \cite{FJK-EPJB00, FJK+EPJB00}.  As far as the first
principles attempts \cite{SUW-PRB95, SUW-PRB97, SU-PMB98, SU-JMMM99,
UZW+PRL99, PKT+PRL00} are concerned, which are currently able to
calculate realistic parameters for a model like Eq.~(\ref{eq:H}), the
technique, presented and applied here, provides a simple and quick
tool to study the finite temperature behavior of thin magnetic films.
As the measurements are performed at finite temperatures, while first
principles calculations usually refer to the ground state, such a
procedure would improve the predictive power of ab--initio theories.
It also should be mentioned, that first attempts to an ab-initio type
description of thin magnetic films at finite temperatures, i.e.
taking into account the coupling of the itinerant nature of the
electrons and the spin degree of freedom, are currently under progress
\cite{razee}.
%
%
\section{Acknowledgements}
This paper resulted from a collaboration partially funded by the RTN
network on 'Computational Magnetoelectronics' (Contract
No. HPRN-CT-2000-00143) and the Research and Technological Cooperation
between Austria and Hungary (OMFB-BMAA, Contract No. A-35/98).
Financial support was provided also by the Center of Computational
Materials Science (Contract No. GZ 45.451/2-III/B/8a), the Austrian
Science Foundation (Contract No. P12146), and the Hungarian National
Science Foundation (Contract No. OTKA T030240 and T029813).
%
%
\section*{Appendix: Derivation of the Curie temperature}

In the high temperature limit ($\beta \rightarrow 0$, $\beta
b_p^{x(z)} \rightarrow 0$) the partition function as given by
Eq. (\ref{eq:part}) can be written up to the first order of the
magnetisation as
\begin{equation}
 Z_p = 2\pi \int\limits_{-1}^1 (1 + \beta b_p^z z)
       e^{\beta\lambda_pz^2}dz 
     = 2\pi \int\limits_{-1}^1 e^{\beta\lambda_pz^2}dz \; .
\end{equation}
Similarly, for the magnetisation in Eq.~(\ref{eq:mfz}) the following
approach can be used
\begin{eqnarray} \nonumber
 m_p^z &=& {2\pi\over Z_p} \int\limits_{-1}^1 z \left[ 1 + \beta
(b_p^z+H)z \right] e^{\beta\lambda_pz^2}dz \\ \label{eq:ht} &=&
{1\over \lambda_p}(b_p^z+H) \left( \dfrac{e^{\beta\lambda_p}}{
\int_{-1}^1e^{\beta\lambda_pz^2}dz } - {1\over 2} \right) \; ,
\end{eqnarray}
where an external magnetic field $H$ has been added to the Hamiltonian
in Eq.~(\ref{eq:mf}).  By substituting the expansion 
\begin{equation}
 {e^{\beta\lambda_p}\over
 \int_{-1}^1e^{\beta\lambda_pz^2}dz } = {1\over 2} + 
 {1\over 3}\beta\lambda_p + {4\over 45} (\beta\lambda_p)^2 + 
 {\mathcal O}(\lambda_p^2) \;
\end{equation}
into Eq. (\ref{eq:ht}) follows
\begin{equation}
 m_p^z = \beta (b_p^z + H) \left( {1 \over 3} + {4 \over 45} \beta
\lambda_p \right) \; .
\end{equation}
Requiring non--vanishing magnetisation at zero external field results
in the following eigenvalue problem
\begin{equation} \label{eq:tc1}
  \sum_{q=1}^N c_{pq}^z \left( 1 + {4\over 15}\beta\lambda_p \right)
  m_q^z = 3k_B T m_p^z \; .
\end{equation}
Let $T_z$ denote the highest value of $T$, for which
Eq.~(\ref{eq:tc1}) is satisfied, i.e. above which no spontaneous
normal--to--plane magnetisation can exist.  A similar procedure can be
applied in order to determine $T_x$, that is, the temperature, at
which the in--plane magnetisation vanishes. Quite obviously, by
neglecting anisotropy effects, for a bulk system the Curie temperature
$T_C=T_x=T_z$ is given by the well--known formula
\begin{equation} \label{eq:tc-bulk}
T_C= {{nJ} \over {3 k_B}} \; , 
\end{equation}
where $n$ denotes the number of nearest neighbors in the bulk.

With exception of very open surfaces such as the BCC(111) surface (see
table \ref{tbl:mad}), the magnetic dipole--dipole coupling constants
$A_{pq}$ fall off exponentially with the distance between layer $p$
and layer $q$.  Therefore, as an approximation we neglect all $A_{pq}$
for $\mid \!\! p \!-\! q \!\! \mid > 1$, which, by recalling the
nearest neighbor approximation for the exchange coupling, implies that
the matrix formed by the elements $c^z_{pq}$ is tridiagonal. The
non--vanishing elements are then written as
\begin{equation}
c_{pp}^z = n_{11}J - A_{11}\omega \quad {\rm and} \quad
c_{p,p-1}^z = c_{p-1,p}^z = n_{12}J - A_{12}\omega \; . 
\end{equation}
Setting $\lambda_p=0$, the solution of the eigenvalue problem
Eq.~(\ref{eq:tc1}) yields
\begin{equation}
T_z^{(0)} = {1 \over {k_B}} \left[ n_{11}J -
A_{11}\omega + 2(n_{12}J - A_{12}\omega) 
\cos \left ({\pi\over N+1} \right) \right]  \quad ,
\end{equation}
with the components of the corresponding normalized eigenvector $u_p =
\sqrt{2 / \left( N+1 \right)}$ $\times\sin \left( p\pi / \left( N+1
\right) \right) $.  Substituting $T_z^{(0)}$ into Eq.~(\ref{eq:tc1})
and using first order perturbation theory with respect to $\lambda_p$,
one gets Eq.~(\ref{eq:tcz}) for $T_z$.  Again, a similar procedure
applies for deriving $T_x$ in Eq.~(\ref{eq:tcx}).
%
%

%
%
%
\vskip 3cm
%
\begin{table}[htbp] \centering 
\begin{tabular}{c|l|l|l|l} \hline\hline
  structure & $A_{11}$ & $A_{12}$ & $A_{13}$ & $A_{14}$ \\ \hline
  SC  (100) & 9.0336   & -0.3275  & -0.00055 & $<$ 10$^{-5}$ \\
  SC  (111) & 11.0342  & 5.9676   & 0.4056   & 0.0146   \\
  BCC (100) & 9.0336   & 4.1764   & -0.32746 & 0.01238  \\
  BCC (111) & 11.0342  & 15.8147  & 5.9676   & -4.0662  \\
  FCC (100) & 9.0336   & 1.4294   & -0.0226  & 0.00026  \\ 
  FCC (111) & 11.0342  & 0.4056   & 0.00113  &  $<$ 10$^{-5}$ \\
  \hline\hline
\end{tabular}
\caption{ 
Dipole--dipole coupling constants as defined in
Eq.~(\protect{\ref{eq:Apq}}) for surfaces of cubic structures.  }
\label{tbl:mad}
\end{table}    
%
%
%
%
%
\
\vskip -2cm
%
%
%
%
%
%
%
%
%

\vskip 3.0cm

%
%
%
%
%
%
%
%
\end{document}